\documentstyle[aps,epsfig,twocolumn,floats]{revtex}

\title{Mechanical response of random heteropolymers}
\author{Phillip L. Geissler and Eugene I. Shakhnovich\\
{\it Department of Chemistry and Chemical Biology,
Harvard University, Cambridge, MA 02138}}

\begin{document}

\wideabs{
\maketitle 

\begin{abstract}
We present an analytical theory for heteropolymer deformation,
as exemplified experimentally by stretching of single protein
molecules.  Using a mean-field replica theory, we determine
phase diagrams for stress-induced unfolding of typical random
sequences.  This transition is sharp in the limit of infinitely
long chain molecules.  But for chain lengths relevant
to biological macromolecules, partially unfolded
conformations prevail over an intermediate range of stress.
These necklace-like structures, comprised of alternating
compact and extended subunits, are stabilized by quenched
variations in the composition of finite chain segments.
The most stable arrangements of these subunits are largely
determined by preferential extension
of segments rich in solvophilic monomers.
This predicted significance of necklace structures explains
recent observations in protein stretching experiments.
We examine the statistical features of select sequences
that give rise to mechanical strength and may thus have
guided the evolution of proteins that carry out 
mechanical functions in living cells.
\end{abstract}
}

\newpage
\section{Introduction}
Several recent experiments have highlighted the mechanical strength 
of proteins involved in muscle elasticity and cell 
adhesion\cite{Fernandez_NAT_98,Fernandez_NAT_99,Fernandez_PNAS_00}.
When pulled from the ends, these
molecules can withstand significant stress 
before their constituent domains unfold from compact native states
to extended coil-like structures.
This stress-induced unfolding occurs sharply,
with threshold forces $f_{\rm t}$ on the order of
100 pN.  In natural units for these systems,
$f_{\rm t} \sim 100 k_{\rm B}T/a$, where
$k_{\rm B}$ is Boltzmann's constant (which we subsequently
set to unity),
$T$ is temperature, and $a\simeq 1$ nm is the size of
a typical amino acid.
By contrast, studies of proteins whose functions are
not mechanical in nature have revealed lower threshold forces
($f_{\rm t} \sim 10 k_{\rm B}T/a$) and less dramatic stretching 
behavior\cite{Bustamante_PNAS_00,Clarke_BIOPHJ_01}.
Specifically, relative fluctuations in restoring force 
are considerably larger than for mechanical proteins, and
the stretching transition is less sharply defined.
Evolution therefore appears to have designed
certain proteins to unfold reproducibly 
under critical stress.

Results of computer simulations have lent details 
to this notion of mechanical design.
Random heteropolymers on a lattice,
which may be viewed as coarse-grained caricatures of
proteins, exhibit stretching behavior
that depends strongly on the sequence of constituent 
monomers\cite{Thir_PNAS_99}.
Typical random sequences elongate
smoothly under stress, passing through one or more long-lived,
partially extended structures.
Sequences selected for their ability to fold rapidly
in the absence of stress, however, undergo a relatively sharp
force-induced transition.
Folding efficiency is thus correlated with mechanical strength
to some extent.
But reaction coordinates for protein folding
are only loosely coupled to the simple mechanical variables
manipulated in stretching experiments\cite{Wolynes_PNAS_99}.
In an ensemble of fast-folding sequences, 
a range of mechanical stabilities is therefore expected, due to
variations in sequence properties that do not affect
folding dynamics.
Native state topology may be one such property, according to
results of protein stretching simulations
with atomistic detail\cite{Schulten_PROT_99,Karplus_PNAS_00}.
But due to the computational expense of such simulations,
they provide only anecdotal insight into the relationship between
sequence and mechanical strength.

A general understanding of heteropolymer deformation can only
be obtained by considering the ensemble of possible sequences.
We have recently described the results of an analytical theory
that takes the diversity of this ensemble properly into 
account\cite{stretching_letter}.
Specifically, the free energies of various conformational states
are averaged over the distribution of sequences, using the replica
trick\cite{Binder_RMP_86}.  In this way, the mechanical response typical of
random heteropolymers is determined.  
This article presents our theoretical approach in detail.

Because we focus on equilibrium free energetics, our theory
directly applies only to {\em reversible} stretching,
i.e., pulling rates that are much slower than 
rates of spontaneous unfolding.
Stretching experiments, on the other hand, have been performed
irreversibly, as evidenced by wide hysteresis\cite{Fernandez_NAT_98}.
Relating theoretical results to these experiments
is made possible by an identity for nonequilibrium dynamics
obtained by Jarzynski\cite{Jarz_PRE_97} 
and by Crooks\cite{Crooks_PRE_00}.
In particular, Hummer and Szabo have shown
that reversible stretching behavior may be
extracted from repeated nonequilibrium measurements\cite{Szabo_PNAS_01}.
Equilibrium results determined in this way
differ only in details from their nonequilibrium counterparts.
For example, the threshold force required to
unfold a mechanical protein reversibly is smaller than the 
corresponding nonequilibrium value,
but the induced transition remains sharp\cite{Szabo_PNAS_01}.
Qualitative predictions of our theory
are thus relevant to current experiments.
Direct comparisons are possible in principle
when experimental measurements have been repeated sufficiently.

\begin{figure}
\centerline{\epsfig{file=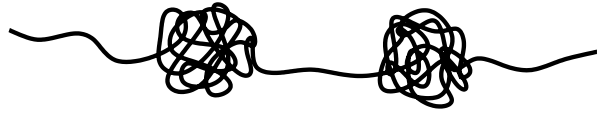,width=8cm}}
\vspace*{0.5cm}
\caption{
Schematic example of a necklace-like polymeric structure.  
Some regions of the chain adopt 
locally compact, globular conformations, while others exist in
extended coil-like states.
}
\label{fig:necklace}
\end{figure}

The coarse features of heteropolymer response do
not differ significantly from those of a homopolymer.
In poor solvent ($T<\theta$), a 
chain molecule is transformed by stress
from collapsed globule to expanded coil.
This transformation occurs abruptly for homopolymers,
as determined by scaling analysis\cite{Zhulina_EL_91}.
Globule deformation is strongly resisted by
the cost of enlarging the polymer-solvent interface, while
a coil is quite pliable.
The ``phase transition'' between these states is thus
accompanied by a sharp change in extension.
At phase coexistence,
the free energetic equivalence of globule and coil
gives rise to necklace-like
structures, in which compact and expanded subunits alternate
within the chain (as sketched in Fig.~\ref{fig:necklace}).
In the case of homopolymers,
these partially extended structures are 
unstable away from coexistence.
A phase diagram for homopolymer stretching is constructed
from this physical picture in Sec.~II.

Necklace structures figure more prominently
in the cases of polyelectrolytes and polyampholytes.
When a significant fraction of monomers carry charge,
fully compact conformations are unstable, 
in analogy to the Rayleigh instability of
charged droplets\cite{Kardar_PRE_95}.  
As a result, the chain segregates into a series of
smaller, tethered globules.
Such necklaces are the ground states of sufficiently
charged polyelectrolytes,
even in the absence of stress.
Stretching these molecules modifies the
the details of structural 
partitioning, reducing the sizes and numbers of compact subunits
and lengthening the string-like subunits that connect them
\cite{Vilgis_EPJE_00,Tamashiro_MM_00}.

Necklaces play an enhanced role in heteropolymer stretching
as well, although for different reasons.
The quenched disorder of random sequences lends
different degrees of mechanical
susceptibility to different regions of the chain.
Depending on the extent of this heterogeneity,
necklace structures may dominate over an appreciable range
of stress.  In effect, the globule-coil coexistence region
is broadened by disorder.
In Sec.~III, we determine the magnitude of this broadening
by analyzing a microscopic model for heteropolymer deformation.
For uncorrelated sequence statistics, we show that necklaces
are stabilized over a stress interval of relative width $N^{-1/2}$,
where $N$ is the number of monomers per molecule.
The effects of correlations within a sequence, also
examined in Sec.~III, suggest that certain statistical patterns are
strongly correlated with mechanical strength.
These patterns are consistent with the 
structures of mechanical proteins designed by evolution.

In seeking a microscopic explanation for the stretching
behavior of proteins, we focus on the effects of
heteropolymeric disorder.  Electrostatic
effects are expected to influence protein stability 
less strongly at physiological 
conditions\cite{Dill_BIOCH_90,Honig_JMB_99}.
We also focus on the response to external stress, rather than
to strain.  Although the extension of protein molecules is 
constrained in experiments, the flexibility of
unfolded segments in these modular structures
mediates the applied strain.  Indeed, provided the contour
lengths of unfolded regions, simple elastic models account for the
measured restoring forces of modular proteins.
Individual, folded domains are thus effectively subjected 
to uniform external stress.
The model we analyze in Sec.~III is tailored to these external
conditions appropriate for experiments and for biological function.

\section{Homopolymer Stretching}
We employ simple, mean-field descriptions of the
conformational states relevant to polymer deformation.
For instance, the free energy of a homopolymer globule
relative to that of an ideal coil, 
\begin{equation}
{\cal F}_{\rm g}(N) \simeq B_0 {\rho} N +
\gamma N^{2/3},
\label{equ:f_globule_homo}
\end{equation}
is dominated by the effective interactions between monomers.
Here, ${\rho}$ is the monomer density, 
and $\gamma$ is the surface tension of the globule-solvent
interface.
The energy density of monomer attractions, $B_0=T-\theta$,
stabilizes the compact state for $T<\theta$.  We focus on 
temperatures below the $\theta$-region, for which the globule
is highly compact, 
i.e., ${\rho}v\sim 1$, where $v$ is
the volume of a monomer.
At this level of description,
the contribution of stress to Eq.~\ref{equ:f_globule_homo}
is negligible within the regime of globule stability.

\begin{figure}
\centerline{\epsfig{file=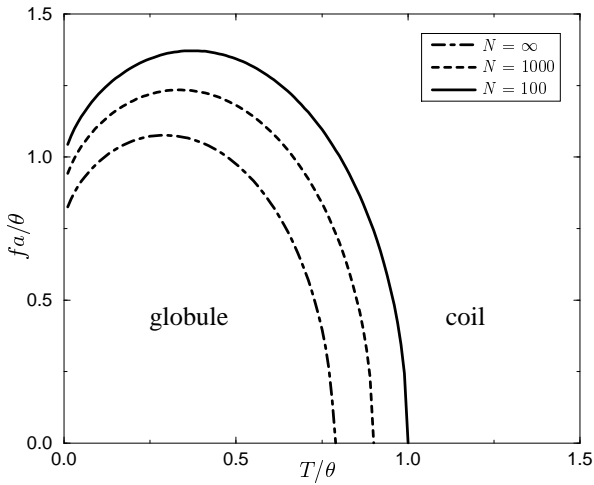,width=8cm}}
\vspace*{1cm}
\caption{
Phase diagram for homopolymer stretching computed from
Eq.~\ref{equ:phase_homo}.  The boundary separating globule
and coil states is plotted 
for chain lengths $N=100$ (dot-dashed curve),
$N=1000$ (dashed curve), and for the infinite chain limit
(solid curve).  Here, we have taken the surface free energy
density of the globule to be comparable to the energy density
of monomer interactions, $\gamma \approx \theta$.
}
\label{fig:homo}
\end{figure}

We represent the coil state by a freely jointed chain
with segment length $a$.  This model provides the
simplest description of polymer flexibility that ensures
a finite maximum chain extension, $Na$.
This condition is important at low temperatures,
where the coil is highly extended under stress.
The free energy of coil deformation is easily 
computed from this model\cite{Grosberg_Khokhlov}:
\begin{equation}
{\cal F}_{\rm c}(N) = -NT\ln{[y(fa/T)]},
\label{equ:f_coil}
\end{equation}
where $y(x) = \sinh{(x)}/x$.
The stretching force at which this extended coil 
coexists with the compact state
is determined by equating ${\cal F}_{\rm c}$ and ${\cal F}_{\rm g}$:
\begin{equation}
f_{\rm t} = y^{-1}\left(\exp{[-(B+\gamma N^{-1/3})/T]}\right) \, T/a.
\label{equ:phase_homo}
\end{equation}
This phase boundary is plotted as a function of temperature
in Fig.~\ref{fig:homo} for various $N$.
Qualitative features of these phase diagrams 
for homopolymer stretching compare well with results of
lattice polymer simulations\cite{Thir_PNAS_99}.
At low temperatures, a reentrant coil phase appears.
This rod-like state involves small fluctuations about 
a fully extended structure, as has been noted in simulation
work\cite{Thir_JPCB_01}.
Similar reentrance has been predicted for the 
mechanical unzipping of DNA at low 
temperatures\cite{Maritan_PRE_01,Shakh_Muk_01}.

\section{Heteropolymer stretching}
\begin{figure}
\centerline{\epsfig{file=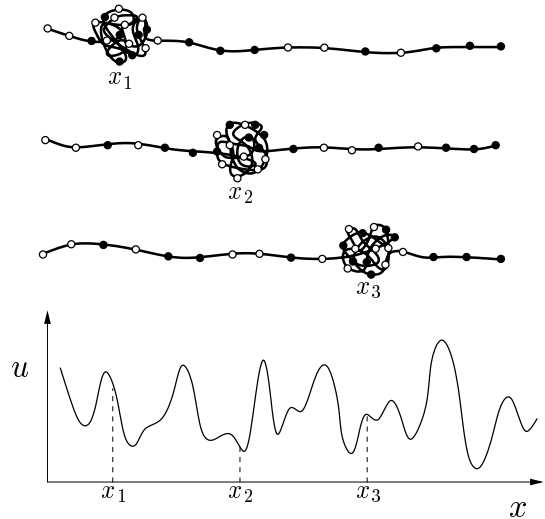,width=7.2cm}}
\vspace*{0.5cm}
\caption{
Random potential $u(x)$ representing the dependence of 
necklace energy on the sequence location $x$ of a globular
subunit.  
Filled and unfilled circles denote
monomers of two different types.
Because the local composition of the sequence differs
at $x_1$, $x_2$, and $x_3$, globular regions at these locations
will have different ground state energies.
More importantly, 
the composition of exposed coil regions
differs for the three cases.
For a random sequence of monomers, the total necklace energy
is thus a random variable for different subunit arrangements.
The statistics of these variations determine the
size of random fluctuations, $\Delta u$.
}
\label{fig:random_potential}
\end{figure}

Heterogeneity of monomer types has several consequences
for the deformation scenario described above.
First, the fully compact state is dominated by
only a few distinct conformations at low temperature.
The corresponding freezing transition has been analyzed 
thoroughly\cite{Shakh_BIOPHCH_89,Shakh_PRE_93}.
For necklace structures, each compact subunit can freeze in this
way.  Because the composition of these subunits is 
randomly distributed,
the details of freezing will differ for each.
Specifically, each subunit will have a 
different ground state energy, and thus a different stability.
In addition,  variations in sequence composition 
will strongly influence the solvation energetics of expanded subunits.
As a result, the susceptibility 
of a given region of the chain to extension
is effectively a random variable.  This situation
is illustrated in Fig.~\ref{fig:random_potential}.

We weigh these effects of disorder 
using a microscopic model of random heteropolymers.
For a particular realization of monomer identities,
$\sigma_i$, the energy of a chain conformation is
\begin{eqnarray}
{\cal H} &=& {\cal H}_0 + \Gamma \sum_{i\in S} 
\sigma_i - {\bf f} \cdot ({\bf r}_N - {\bf r}_1),
\label{equ:h_hetero}
\\
{\cal H}_0 &=& \sum_{i,j=1}^N \delta({\bf r}_i - {\bf r}_j)
(B_0 + \chi \sigma_i \sigma_j).
\label{equ:h_0} 
\end{eqnarray}
Here, ${\bf r}_i$ denotes the position of the $i$th monomer
in the chain, and ${\bf f}$ is the external stretching force
coupled to the end-to-end vector, ${\bf r}_N - {\bf r}_1$.
The connectivity of consecutive monomers is implicit
in Eq.~\ref{equ:h_hetero}.  We take the links between
connected monomers 
$i$ and $i+1$ to be distributed
according to a function $g(|{\bf r}_{i+1}-{\bf r}_{i}|)$
with range $a$.
Unconnected monomers $i$ and $j$ 
interact only when they are in contact,
as described by the Dirac delta function in Eq.~\ref{equ:h_0}.
The heteropolymeric part of this interaction
depends on the identities of the monomers involved.
Two monomer types, $\sigma_i=\pm1$, are possible
at each point in the sequence.  These types could
represent, for example, amino acids with hydrophilic and hydrophobic
side chains.
We choose $\chi<0$, so that each monomer attracts others
of the same type most strongly.
The sequence of monomer types $\sigma_i$ is fixed for each
realization of the heteropolymer.  
As in related problems, this quenched disorder
requires careful treatment.

The second term in Eq.~\ref{equ:h_hetero}
describes the solvation energetics of monomers
that are exposed to the external environment.  The sum extends
only over the set $S$ of exposed monomers.  Depending
implicitly on chain conformation, this term mimics
a many-body aspect of the hydrophobic effect,
the tendency to bury unfavorably solvated regions of a solute.
We take $\Gamma>0$, so that 
monomers of type $\sigma_i=-1$ are preferentially solvated,
In the following sections, we analyze the energetics of
Eqs.~\ref{equ:h_hetero} and~\ref{equ:h_0} for the
conformational states important to mechanical response.
With these results, we construct
phase diagrams for heteropolymer stretching.

\subsection{Globule}
The first term in Eq.~\ref{equ:h_hetero},
${\cal H}_0$, 
defines a model of a random copolymer in the absence of explicit
solvation energetics ($\Gamma=0$) and stretching force (${\bf f}=0$).
A mean field theory for the globular state of this 
model was analyzed in Ref.~\cite{Shakh_PRE_93}
using the replica trick.
At a critical temperature, $T_{\rm c}$, the heteropolymer freezes 
into $O(1)$ low energy conformations.
This transition is accompanied by 
one-step replica symmetry breaking, indicating that  
the hierarchy of basins of attraction in conformation space
is one level deep.
This result is consistent with a random energy model
of the system.  In other words, the density of states
is well represented by drawing energies at random 
from a Gaussian distribution, $P(E)$, with appropriate
mean, $\overline{E}$, and variance, $\Delta$.
Our analysis of the globular state for the
model defined by Eq.~\ref{equ:h_hetero}
(with nonzero $\Gamma$ and ${\bf f}$) closely follows
that of Ref.~\cite{Shakh_PRE_93}.  We focus on 
the way in which solvation modifies the effective
distribution of energies.  As in the homopolymeric case,
we neglect the effect of stretching force on
globule free energetics.

In order to compute properties characteristic
of an ensemble of random heteropolymers, we must 
average over their sequences.  
Because the disordered sequence is quenched,
the average is
correctly performed on the logarithm of the partition function,
$Z$, rather than on $Z$ itself.
This mathematically awkward procedure
can be accomplished using the replica trick\cite{Binder_RMP_86},
\begin{equation}
\langle \ln{Z} \rangle_{\rm av} = 
\lim_{n\rightarrow 0} \frac{\langle Z^n\rangle_{\rm av} - 1}{n},
\label{equ:trick}
\end{equation}
where $\langle \ldots \rangle_{\rm av}$ denotes an average
over realizations of the random sequence.
For integer values of
$n$, the quantity $\langle Z^n\rangle_{\rm av}$
has the form of a partition function for $n$
coupled replicas of the original system.
For the model of Eq.~\ref{equ:h_hetero},
\begin{eqnarray}
\langle Z^n \rangle_{\rm av} &&= 
\left\langle
\int
\left[ \prod_{\alpha=1}^n \prod_{j=1}^N d{\bf r}_j^\alpha 
g({\bf r}_{j+1}^\alpha - {\bf r}_{j}^\alpha) \right]
\right.
\nonumber
\\
\times &&
\left.
\exp{\left[-{B_0 \over T} \sum_{\alpha=1}^n \sum_{i,j=1}^N 
\delta({\bf r}_{i}^\alpha-{\bf r}_{j}^\alpha) \right]}
\right.
\nonumber
\\
\times && \left.
\exp{\left[-\sum_{i,j} {\chi\over T}\sigma_i 
\delta({\bf r}_{i}^\alpha-{\bf r}_{j}^\alpha)\sigma_j 
-{\Gamma \over T}\sum_{i\in S} \sigma_i \right]}
\right\rangle_{\rm av},
\label{equ:Zn}
\end{eqnarray}
where ${\bf r}_i^{\alpha}$ is the position of the $i$th
monomer of replica number $\alpha$.
In order to perform the average in Eq.~\ref{equ:Zn}, 
we first rewrite the right hand side as
\newpage
\begin{eqnarray}
&&
\left\langle
\int {\cal D}{\bf r}_j^\alpha 
g({\bf r}_{j+1}^\alpha - {\bf r}_{j}^\alpha)
\exp{[-{B_0 \over T}\sum_{\alpha=1}^n \sum_{i,j=1}^N 
\delta({\bf r}_{i}^\alpha-{\bf r}_{j}^\alpha) ]}
\right.
\nonumber
\\
&&
\times
\exp{[b\sum_\alpha\int d{\bf R} \sum_i \sigma_i 
\delta({\bf r}_i^\alpha-{\bf R})
\sum_j \sigma_j 
\delta({\bf r}_j^\alpha-{\bf R})
}
\nonumber
\\
&&
\left.
+c\sum_\alpha\int d{\bf R} 
\rho_{{\rm s},\alpha}({\bf R}) \sum_i \sigma_i 
\delta({\bf r}_i^\alpha-{\bf R})
]
\right\rangle_{\rm av},
\label{equ:rhs}
\end{eqnarray}
where $b=-\chi/T$, $c=-\Gamma/T$, and
$\int {\cal D}{\bf r}_j^\alpha$ denotes an
integration over the monomer positions of all replicas.
We have also defined 
\begin{equation}
\rho_{{\rm s},\alpha}({\bf R}) = \sum_{i\in S} 
\tilde{\delta}({\bf R}-{\bf r}_i^{\alpha})
\end{equation}
as the density of
exposed monomers at position ${\bf R}$, i.e., the spatial
pattern formed by the globule boundary.
Here, $\tilde{\delta}({\bf r})$ is $1$ if ${\bf r}=0$,
and vanishes otherwise.
We perform a Hubbard-Stratonovich transformation
with respect to the field
\begin{equation}
\sum_i \sigma_i \delta({\bf r}_i^{\alpha}-{\bf R}),
\end{equation}
by introducing a conjugate field $\psi_{\alpha}({\bf R})$.
With this transformation, the second exponential in
Eq.~\ref{equ:rhs} becomes
\begin{eqnarray}
\int &&
d{\cal D}\psi_\alpha({\bf R}) \exp{\left[-{1\over 4b}
\sum_\alpha \int d{\bf R} \psi_\alpha^2({\bf R})\right]}
\nonumber
\\
&&
\times
\exp{\left[
\sum_\alpha \int d{\bf R} \psi_\alpha({\bf R}) \sum_i \sigma_i 
\delta({\bf r}_i^\alpha -{\bf R})\right]}
\nonumber
\\
&& \times
\exp{\left[
{c\over 2b} \sum_\alpha \int d{\bf R} \psi_\alpha({\bf R}) 
\rho_{{\rm s},\alpha}({\bf R})
\right.}
\nonumber
\\
&&
\quad\quad\quad\quad
\left.
- {c^2\over 4b} \sum_\alpha \int d{\bf R} \rho_{{\rm s},\alpha}^2({\bf R})
\right].
\end{eqnarray}

The average over sequence realizations may now be easily performed,
yielding an action that is highly nonlinear in $\psi_{\alpha}({\bf R})$.
As was done in Ref.\cite{Shakh_PRE_93}, we retain only the first term
in a high-temperature expansion of the nonlinearity.
As in that work, inclusion of additional terms does not change the leading 
order behavior of relevant order parameters.  This simplification
corresponds to treating monomer types as Gaussian, rather than binary,
variables, with distribution
\begin{equation}
w(\sigma_i) = (2\pi \mu^2)^{-1/2} \exp{(-\sigma_i^2/2\mu^2)}.
\label{equ:mondist}
\end{equation}
The variety of interaction strengths 
described by Eq.~\ref{equ:mondist}
might be appropriate for monomers that 
come into contact with a variety of relative orientations.
It could also describe a heteropolymer
with more than two possible monomer types.

Averaging over the effective distribution of monomer identities,
and scaling the field $\psi_{\alpha}({\bf R})$ by $2b$,
we obtain
\begin{eqnarray}
&&
\langle Z^n \rangle_{\rm av} =
\nonumber
\\
&&
\quad
\left\langle
\exp{\left[-{B_0\over T}\sum_\alpha \int d{\bf R} \rho_\alpha^2({\bf R})
- {c^2\over 4b} \sum_\alpha 
\int d{\bf R} \rho_{{\rm s},\alpha}^2({\bf R})\right]}
\right.
\nonumber
\\
&& \quad
\times
\int {\cal D}\psi_\alpha({\bf R})
\exp{\left[-b\int d{\bf R}\psi_\alpha^2({\bf R})
\right.}
\nonumber
\\
&&
\quad
\left.
+2 b^2 \mu^2 \sum_{\alpha,\beta} \int d{\bf R}_1 d{\bf R}_2
\psi_\alpha({\bf R}_1)\psi_\beta({\bf R}_2) 
Q_{\alpha\beta}({\bf R}_1,{\bf R}_2)
\right]
\nonumber
\\
&& \quad
\left. \times
\exp{\left[c \sum_\alpha \int d{\bf R} 
\psi_\alpha({\bf R}) \rho_{{\rm s},\alpha}(\bf R)
\right]}
\right\rangle_{\rm th},
\label{equ:Znav}
\end{eqnarray}
where $\langle \ldots \rangle_{\rm th}$ denotes a thermal 
average over the statistics of monomer links imposed
by $g({\bf r}_{i+1}^\alpha - {\bf r}_{i}^\alpha)$.
In Eq.~\ref{equ:Znav}, we have additionally
introduced two fields:
a single-replica density field,
$\rho_\alpha({\bf R} = \sum_i \delta({\bf r}_i -{\bf R})$,
and a field describing the conformational similarity of
two replicas,
\begin{equation}
Q_{\alpha \beta}({\bf R}_1,{\bf R}_2) = \sum_i 
\delta({\bf r}_i^{\alpha}-{\bf R}_1) 
\delta({\bf r}_i^{\beta}-{\bf R}_2).
\end{equation}
Since density fluctuations are negligible in the globular state,
$\rho_{\alpha}({\bf R})$ may be approximately replaced by its
mean value, $\rho \sim v^{-1}$, where $v$ is the volume of a
typical monomer.
Similarly, the surface density, $\rho_{\rm s}({\bf R})$,
is essentially fixed in the globular state.
The replica overlap function, $Q_{\alpha \beta}({\bf R}_1,{\bf R}_2)$, 
however, is an important measure of the population of different
conformational basins of attraction. It is thus a useful
order parameter to describe freezing of random
heteropolymers into their lowest energy conformations.

Because density is virtually constant within the globule,
$Q_{\alpha \beta}$ is a function of ${\bf R}_1-{\bf R}_2$
only, and its Fourier transform depends on a single wavevector
${\bf k}$:
\begin{equation}
\hat{Q}_{\alpha \beta}({\bf k}) = \int d({\bf R}_1-{\bf R}_2)
Q_{\alpha \beta}({\bf R}_1,{\bf R}_2) \exp{[i{\bf k}\cdot
({\bf R}_1-{\bf R}_2)]}.
\end{equation}
With a Fourier representation of $\psi_{\alpha}({\bf R})$,
the right hand side of Eq.~\ref{equ:Znav} becomes
\begin{eqnarray}
&&
\left\langle
\int {\cal D}\hat{\psi}_{\alpha}({\bf k})
\exp{\left[
V \sum_{\alpha,\beta} \sum_{\bf k} P_{\alpha \beta}({\bf k})
\hat{\psi}_{\alpha}({\bf k})\hat{\psi}_{\beta}(-{\bf k})
\right.}
\right.
\nonumber
\\
&&
\qquad\qquad
\left. \left.
+V c \sum_{\alpha}\sum_{\bf k} \hat{\rho}_{\rm s}({\bf k})
\hat{\psi}_{\alpha}(-{\bf k})
\right]
\right\rangle_{\rm th},
\label{equ:FT}
\end{eqnarray}
where
\begin{equation}
P_{\alpha \beta}({\bf k}) = 
-b \delta_{\alpha \beta} + 2 \mu^2 b^2 \hat{Q}_{\alpha \beta}({\bf k}).
\end{equation}
In Eq.~\ref{equ:FT} we have omitted the first exponential of
Eq.~\ref{equ:Znav}, which contributes
an irrelevant multiplicative constant.

Following the analysis of Ref.~\cite{Shakh_PRE_93},
we use the replica overlap function 
as the order parameter for a mean field
theory of heteropolymer freezing.
To this end, we rewrite Eq.~\ref{equ:FT} as a functional
integral over possible realizations of $Q_{\alpha \beta}$:
\begin{equation}
\langle Z^n \rangle_{\rm av} = \int 
{\cal D}\hat{Q}_{\alpha \beta}({\bf k})
\exp{[-E\{\hat{Q}_{\alpha \beta}({\bf k})\}
+S\{\hat{Q}_{\alpha \beta}({\bf k})\}
]}
\label{equ:MFT}
\end{equation}
Here, $E$ is the effective energy of a particular realization:
\begin{eqnarray}
&&
E\{\hat{Q}_{\alpha \beta}({\bf k})\} = \ln{
\int {\cal D} \psi_{\alpha}({\bf R}) 
}
\nonumber
\\
&& 
\qquad\qquad
\times
\exp{\left[
V \sum_{\alpha,\beta} \sum_{\bf k} P_{\alpha \beta}({\bf k})
\hat{\psi}_{\alpha}({\bf k})\hat{\psi}_{\beta}(-{\bf k})
\right.}
\nonumber
\\
&& 
\qquad\qquad
\left.
+V c \sum_{\alpha}\sum_{\bf k} \hat{\rho}_{\rm s}({\bf k})
\hat{\psi}_{\alpha}(-{\bf k})
\right]
\\
&&
= \int d{\bf k} \left[
\ln{\det P_{\alpha \beta}({\bf k})}
-{c^2\over 4} \sum_{\alpha, \beta}
|\hat{\rho}_{\rm s}({\bf k})|^2
P^{-1}_{\alpha \beta}({\bf k})
\right].
\label{equ:E}
\end{eqnarray}
Similarly, $S$ is an effective entropy describing the number
of conformations consistent with a particular realization
of $Q_{\alpha \beta}$:
\begin{eqnarray}
&&
S\{\hat{Q}_{\alpha \beta}({\bf k})\} = 
\nonumber
\\
&& \qquad
\ln{
\left\langle
\delta\left(
\hat{Q}_{\alpha \beta}({\bf k}) - 
V \sum_i \exp{[i {\bf k}\cdot 
({\bf r}_i^{\alpha}-{\bf r}_i^{\beta})
]}\right)
\right\rangle_{\rm th}
}.
\label{equ:S}
\end{eqnarray}
We will approximate the integral in Eq.~\ref{equ:MFT} by
optimizing the free energy with respect to 
replica overlap.

We imagine that a hierarchy exists for
basins of attraction in conformation space,
as is done in the theory of spin glasses\cite{Binder_RMP_86}.
It is then natural to sort replicas into groups, such that replicas
belonging to the same group overlap most strongly
This grouping determines the structure of $Q_{\alpha \beta}$.
If $\alpha$ and $\beta$ are in the same group, 
$Q_{\alpha \beta}({\bf R}_1,{\bf R}_2)$ is nearly 
$\rho \delta({\bf R}_1-{\bf R}_2)$.  If $\alpha$ and $\beta$
belong to widely different groups, 
$Q_{\alpha \beta}({\bf R}_1,{\bf R}_2)\approx 0$.
The one-step replica symmetry breaking demonstrated in
Ref.~\cite{Shakh_PRE_93} is not altered by the solvation
energetics in Eq.~\ref{equ:h_hetero},
because the scaling properties of $Q_{\alpha \beta}$
are unchanged.
Consequently, overlap between replicas is binary:
\begin{equation}
\hat{Q}_{\alpha \beta}({\bf k}) = 
\cases{
\rho, & for $\alpha, \beta$ in the same group, \cr
0, & for $\alpha, \beta$ in different groups.
}
\label{equ:Q}
\end{equation}
Note that $\hat{Q}_{\alpha \beta}({\bf k})$ is independent of ${\bf k}$,
since replica overlap is either absent or microscopically complete.
Together with the number of replicas in each group,
Eq.~\ref{equ:Q} specifies $\hat{Q}_{\alpha \beta}({\bf k})$ 
completely.

The limit $n\rightarrow 0$ in Eq.~\ref{equ:trick}
is most conveniently taken using a continuous representation of the 
replica overlap matrix.  In this limit, the ``number'' of replicas
in a group, $x_0$, lies between 0 and 1, and summations over replica
indices are replaced by integrations on the interval $[0,1]$.
The first term in Eq.~\ref{equ:E} was computed in 
Ref.~\cite{Shakh_PRE_93} using the continuous form of 
$Q_{\alpha \beta}$ introduced by Parisi:
\begin{equation}
\ln{[\det{P_{\alpha \beta}}]} = 
\ln{b} + \frac{\ln(1-\gamma x_0)}{x_0},
\end{equation}
where $\gamma = 2 b \mu^2 \rho$.
We evaluate the second term in Eq.~\ref{equ:E} using
identities derived in Ref.~\cite{Parisi_JPP_91}
for the Parisi matrix:
\begin{equation}
\sum_{\alpha,\beta} |\hat{\rho}_{\rm s}({\bf k})|^2 
P^{-1}_{\alpha\beta}({\bf k}) = 
\frac{n}{b(1-\gamma x_0)} A,
\end{equation}
where $A$ is the surface area of the globule.
The loss of entropy due to the grouping of replicas described
by $Q_{\alpha \beta}$ was also computed in Ref.~\cite{Shakh_PRE_93}
as $S = {N n S / x_0}$. 
Here, $s=\ln a^3/v$ is the entropy loss per monomer
of constraining
a replica to correspond to other replicas 
in its group at a microscopic level.
Combining these results, and noting that wavevector summations
contribute only unimportant factors of volume, we obtain the
replica free energy density:
\begin{equation}
{{\cal F}(x_0)\over nN} = \ln{b} + {\ln{(1-\gamma x_0)}\over x_0}
-{c^2\over 4b}(1-\gamma x_0)^{-1}{A\over N}
-{s\over x_0}.
\label{equ:f_rep}
\end{equation}
Eq.~\ref{equ:f_rep} differs from the corresponding result
in Ref.\cite{Shakh_PRE_93} only by the term proportional
to $A/N \sim N^{-1/3}$.

We now employ a mean field approximation
by optimizing the free energy density with respect to $x_0$.
(Because the number of pairs of replicas is negative in the
limit $n\rightarrow 0$, the appropriate extremum is in fact a
maximum of $F(x_0)$.)  To lowest nonvanishing order in $x_0$,
the mean field solution is
\begin{equation}
x_0 = 
\gamma^{-1}\sqrt{ \frac{2s}
{1+ {c^2 A/ 2 b \gamma N}} }.
\label{equ:x0}
\end{equation}
From Eq.~\ref{equ:x0}, 
we may identify the transition temperature, $T_{\rm c}$,
may be identified at which freezing occurs, i.e., at which
$x_0$ first deviates from unity:
\begin{equation}
T_{\rm c} = (2 s)^{-1/2}
(-2\chi \mu^2 \rho - {\Gamma^2 A \over 4 \chi N})
+O(N^{-2/3}).
\label{equ:Tc}
\end{equation}
Comparing this result with Eq.~3.9 of Ref.~\cite{Shakh_PRE_93},
we find that the solvation term in Eq.~\ref{equ:h_hetero} raises
$T_{\rm c}$ by an amount $\sim N^{-1/3}$.

Together with the average energy of non-native conformations,
the freezing temperature in Eq.~\ref{equ:Tc} determines 
the parameters of a random energy model 
corresponding to the random heteropolymer.
The effective distribution of energies is given by
\begin{equation}
P(E) = (2\pi)^{1/2} \exp{[-(E-\overline{E})^2/N\Delta^2]},
\end{equation}
where $\Delta = \sqrt{2s}T_{\rm c}$
and $\overline{E}=B_0 \rho N$.
This distribution is dominated by
states in the interval
$\overline{E}-N^{1/2}\Delta < E < 
\overline{E}+N^{1/2}\Delta$.
At energies just below a critical value,
$E^*=\overline{E} - N\Delta s^{1/2}$,
the number of states is $O(1)$, while just above $E^*$
the number is  exponentially large.
The ground states of particular random sequences
are distributed narrowly about $E^*$\cite{Derrida_PRB_81}.
Solvation thus lowers
the average ground state energy by an amount 
$(\Gamma^2 {\rho}/ 4 |\chi|)
s^{1/2} A$.

\begin{figure}
\centerline{\epsfig{file=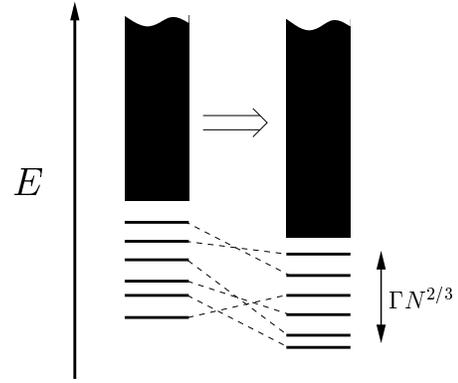,width=6cm}}
\vspace*{1cm}
\caption{
Schematic effect of solvation on the distribution of
globule energies.
Low-lying energy levels, 
as well as the essentially continuous part
of the energy spectrum,
are sketched at left for a heteropolymer without differential
solvation of monomer types.
Surface energetics reorder these levels on a scale
$\Gamma N^{2/3}$ (as shown at right).  The basic form of the
distribution is unchanged.
}
\label{fig:rem}
\end{figure}

Solvation of the globule surface selects a ground
state from the set of conformations with monomer
interaction energies $E< E^* + \Gamma N^{2/3}$.
This selection is illustrated in Fig.~\ref{fig:rem}.
If the energy scale of solvation is small,
$|\Gamma| \ll |\chi|$, the shift in ground state
energy will be a negligible fraction of the optimum
surface energy, $\Gamma \rho A$.
In this case, the set of low-energy conformations
from which to select is small, and it is unlikely that 
one of these conformations presents a predominantly
solvophilic surface.  If, on the other hand,
$|\Gamma| \approx |\chi|$, solvation can be an
important factor in determining the ground state.
In this case, there is a reasonable probability that
a conformation with $E< E^* + \Gamma N^{2/3}$ has favorable
solvation energy.  Here, the shift in ground state energy
will be comparable to $\Gamma \rho A$, and the surface of 
the native state will be largely solvophilic.  This solvation
effect does not strongly influence the freezing behavior
studied in Ref.\cite{Shakh_PRE_93}.  But it does represent
the energetic contribution most sensitive to variations in sequence
composition.  It is therefore significant for our analysis of
necklace structures.

\subsection{Coil}
For the coil state of a heteropolymer,
we must add solvation energetics
to the free energy in Eq.~\ref{equ:f_coil}.
For simplicity, we consider only random sequences whose
total composition is fixed by
$\sum_i \sigma_i = 0$.
Since nearly all monomers are exposed to solvent in the 
expanded coil state, this constraint causes the solvation
energy in Eq.~\ref{equ:h_hetero} to vanish.
Thus, the free energy of this heteropolymeric coil
is identical to that of the homopolymeric coil discussed
in Sec.~II.
In the next section, we consider necklace structures
of a random heteropolymer.  In these structures, short
segments of the chain exist in coil-like states.
The sequence composition of these exposed segments
is not constrained, and the solvation energy need not vanish.
We will show this to be an important stabilizing effect
for necklace structures.

\subsection{Necklace structures}
For a random heteropolymer,
the numbers and sizes of globular subunits
do not provide an adequate description of a
necklace-like structure.
Due to the heterogeneity of monomer interactions,
the energy of such a structure depends as well on the
arrangement of subunits within the sequence.
This dependence on globule location, $x$,
can be represented as the effect of an external, random potential
$u(x)$ with fluctuations of size $\Delta u$,
as illustrated in Fig.~\ref{fig:random_potential}.
If $\Delta u$ is large,
it is likely that a particular arrangement lies much lower
in energy than others.
Over some range of stretching forces, it is
possible that such low-energy necklace structures are
preferred over pure globule and coil states.
In this case, necklaces will play
a significant role in the response to stretching.
Our analysis of heteropolymeric necklaces is guided
by this perspective.
First, we characterize the statistics of the effective random
potential.  We then identify situations in which quenched disorder
stabilizes necklace structures significantly.

Although we take sequence composition to be fixed
for the chain as a whole, the composition $q$ of local regions
is distributed according to
\begin{eqnarray}
p(q) \propto 
\exp{\left[-{1\over 2}q^2 ({1\over M}-{1\over N})^{-1}\right]},
\label{equ:qdist}
\\
q = {1\over M} \sum_{i\in {\rm segment}} \sigma_i,
\end{eqnarray}
where $M$ is the length of the segment.
Because different segments of the sequence have different compositions,
their local free energetics will vary.
In a region with composition $q$, the apparent distribution of 
monomer types is modified from Eq.~\ref{equ:mondist},
\begin{equation}
w(\sigma_i;q) \propto \exp{[-(\sigma_i-q)^2/2(1-q^2)\mu^2]}.
\end{equation}
These modified sequence statistics alter the local distribution
of energies, for both globule and coil subunits.

In the context of the random energy model discussed in Sec.~III.A,
the variation in local sequence statistics modulates the mean and 
variance of conformational energies.
In effect, each segment of the sequence has a different
associated random energy model. 
A globular subunit will therefore have a different ground
state energy for each location in the sequence.
Specifically, the local value of $q$ shifts the mean energy
by an amount $\chi q^2 \rho M + \Gamma q \rho A$,
and reduces the variance of monomer identities, $\mu$,
by a factor $\sqrt{1-q^2}$.
As a result, the characteristic
ground state energy in a sequence region
with composition $q$ is
\begin{eqnarray}
&&
E^* = B_0 \rho M + \chi q^2 \rho M + \Gamma q \rho A
\nonumber
\\
&& \qquad\qquad\qquad\qquad
+\chi \mu^2 (1-q^2)\rho M 
+ O(M^{1/3}).
\end{eqnarray}
Variations in $E^*$ along the sequence contribute to the
random potential $u(x)$.
The magnitude of this contribution is computed by averaging
variations in $E^*$ over the distribution of $q$ in 
Eq.~\ref{equ:qdist}:
\begin{equation}
\sqrt{\langle  (\delta E^*)^2\rangle_q} \sim \Gamma \mu \rho M^{1/6}.
\end{equation}
Fluctuations in $u(x)$ due to energetics of a globular
region of length $M$ thus arise from solvation at leading order,
and grow rather slowly with increasing $M$.

Variations in the solvation of coil regions are more sizable.
In a region of length $M$ and composition $q$, the solvation
energy is $\Gamma q M$.
Fluctuations of this energy are of size
$\Gamma \mu M^{1/2} (1-M/N)^{-1/2}$.
These variations are considerably larger
than those of globule energetics for large $M$,
and set the scale of $\Delta u$. 
We show below that this solvation effect is sufficient to
stabilize necklace structures for long but finite chains.

We focus on necklace structures including
$m$ globular regions, each with $M$ monomers.
Thermodynamics of this class of structures 
may be approximated by
drawing $\Omega$ values at random from a Gaussian
distribution with variance 
$\Delta u$.
Here, $\Omega$ is the number of
statistically independent arrangements of the globular
regions, $\Omega\approx [M^{-m}(N-M+1)(N-2M+1)\ldots(N-mM+1)/m!]$.
The thermodynamics of large systems with
independently distributed random energies
are well known\cite{Derrida_PRB_81,Derrida_PRL_80}.
In this case, the free energy of spontaneous fluctuations
in the random potential $u(x)$ is given by
\begin{equation}
{\cal F}_{\rm rand}(M,m) = \cases{
-\ln{\Omega} \,\,T[1+({T_{\rm \ell} / T})^2], & $T>T_{\rm \ell}$\cr
-2\ln{\Omega}\,\,T_{\rm \ell}, & $T\leq T_{\rm \ell}.$\cr}
\label{equ:fneck}
\end{equation}
In Eq.~\ref{equ:fneck}, 
$T_{\ell}=\Gamma \mu (mM)^{1/2} (1-mM/N)^{1/2}/
(2\ln{\Omega})^{1/2}$ 
is the temperature at
which globular regions become localized in the sequence.
For $T<T_{\ell}$, the necklace has a frozen arrangement
of subunits, and does not reorganize.
Combining Eq.~\ref{equ:fneck} with Eqs.~\ref{equ:f_globule_homo} 
and~\ref{equ:f_coil}, we obtain the total free energy
of a necklace structure,
\begin{equation}
{\cal F}_{\rm neck}(M,m) = m {\cal F}_{\rm g}(M) + 
{\cal F}_{\rm c}(N-mM)
+{\cal F}_{\rm rand}(M,m).
\label{equ:ftotal}
\end{equation}
The pure globule and coil states of a heteropolymer are also described
by Eq.~\ref{equ:ftotal} for $M=N$ and $M=0$, respectively.

\begin{figure}
\centerline{\epsfig{file=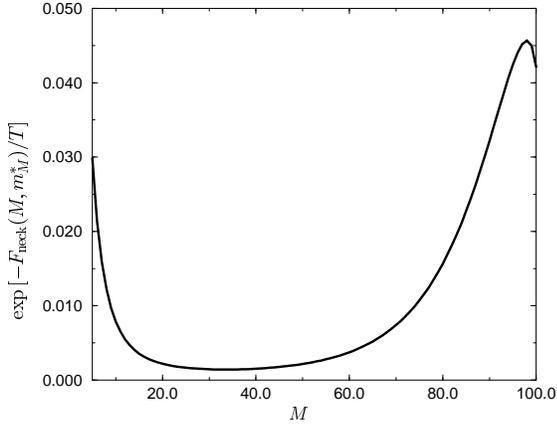,width=8cm}}
\vspace*{0.2cm}
\caption{
Distribution of necklace structures, $p(M)$,
for the thermodynamic state $T=0.5 \theta$, $f=0.95 \theta/a$
with $N=100$.  
Necklaces consisting of a single large globule ($M=O(N)$)
and short coil regions are most probable at these conditions.
But necklaces with many small globules ($M\ll N$) are also
strongly represented.
Globules of intermediate size ($M\approx N/2$) occur with
a small but nonnegligible probability.
A minimum size of 5 monomers is chosen for globular
regions.
}
\label{fig:neckdist}
\end{figure}

The scaling of Eq.~\ref{equ:fneck} suggests that
the distribution of globular region sizes may be quite broad.
For a necklace with
small globules ($M\ll N$), ${\cal F}_{\rm rand}$ is optimized
with a large number of regions, $m=O(N)$. 
In this case, ${\cal F}_{\rm rand}\sim N^{1/2}$.
For large globules, the constraint that $mM<N$ requires
that $m=O(1)$.  In this case as well, 
${\cal F}_{\rm rand}\sim N^{1/2}$.
Surface effects will yield a preference for
a small number of large globular regions,
but this scaling indicates that a broad ensemble of globule sizes
may be important at a single thermodynamic state.
This possibility is demonstrated in Fig.~\ref{fig:neckdist},
in which the distribution of globule sizes,
$p(M)\propto \exp{[-\sum_m'
{\cal F}_{\rm neck}(M,m)/T]}$, is plotted as a function of
$M$ for a thermodynamic state near the globule-coil transition.
Here, primed sums are restricted to $m \le N/M$.
The weight of necklaces consisting of a single large globule
is comparable to that of necklaces comprised of many small globules.
As a result, fluctuations in polymer extension are
considerable near the transition.

\begin{figure}
\centerline{\epsfig{file=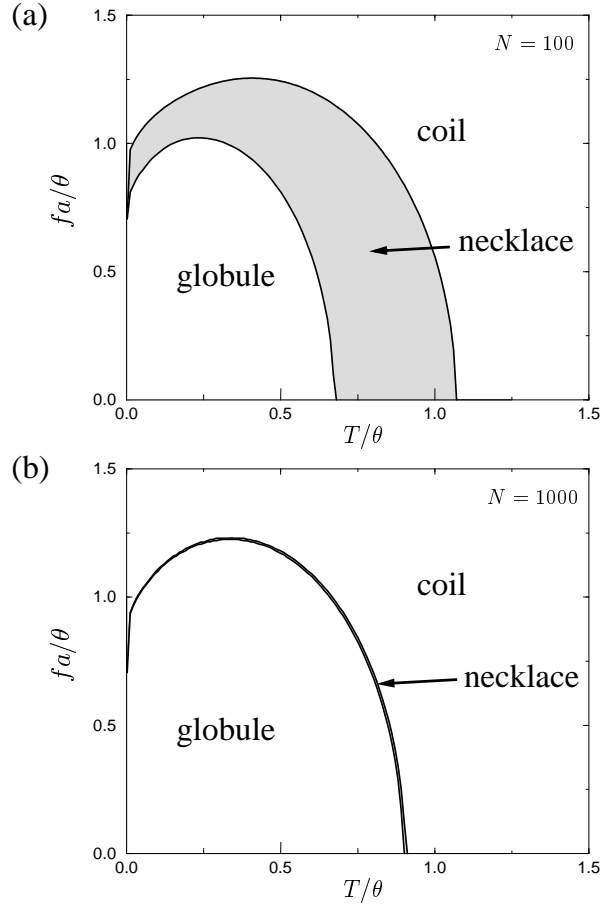,width=8cm}}
\vspace*{0.1cm}
\caption{
Phase diagrams for random heteropolymer stretching
computed from Eq.~\ref{equ:phig}.
Shaded regions mark a necklace phase,
in which the fraction of monomers belonging to 
globular regions lies between $0.05$ and $0.95$.
For these calculations, we have chosen the scale of
monomer solvation energetics to coincide with that 
of interactions between monomers, 
$\Gamma \approx \theta$.
We have also chosen the variance of monomer identities,
$\mu$, to be unity.
}
\label{fig:hetero}
\end{figure}

We determine a phase diagram for stretching of random heteropolymers 
by computing the total fraction of monomers belonging to globular
regions, $\phi_{\rm g}$:
\begin{equation}
\phi_{\rm g} \simeq \frac{\sum_M \sum_m' mM 
\exp{[-{\cal F}_{\rm neck}(m,M)/T]}}
{\sum_M \sum_m'\exp{[-{\cal F}_{\rm neck}(m,M)/T]}}.
\label{equ:phig}
\end{equation}
When $\phi_{\rm g}>0.05$, we consider the heteropolymer to be 
in a globular state.  Similarly, when $\phi_{\rm g}<0.95$,
we consider it to be in a coil state.
In the intermediate regime, $0.05<\phi_{\rm g}<0.95$, 
necklace structures are prevalent.  Results are plotted in 
Fig.~\ref{fig:hetero}.  
For $N=100$, the extent of the necklace phase
prevents the possibility of a sharp
transition from globule to coil.
This result is consistent with simulated stretching
of a short ($N=27$) random heteropolymer, suggesting that
intermediates states described in Ref.~\cite{Thir_PNAS_99}
should be identified with
the necklace structures we consider here.
But the stabilization of necklaces arising from
${\cal F}_{\rm rand}$ scales only as $N^{1/2}$, 
and is overcome for long chains by
$O(N)$ contributions of the first two terms in Eq.~\ref{equ:fneck}.
The range of stretching forces over which necklaces are stable
is therefore significantly 
diminished for $N=1000$, and vanishes as $N\rightarrow \infty$.
Stretching behavior of infinitely long random heteropolymers is
indistinguishable from that of homopolymers.
But for chain lengths relevant to macromolecules,
necklace structures can play an important role.
The large fluctuations in extension accompanying 
these structures
explains the observed stretching behavior of proteins
like barnase\cite{Clarke_BIOPHJ_01} that are not designed
for mechanical functions.

\subsection{Sequence statistics}
In the previous section we showed that the distribution
of local sequence compositions plays an important role
in the stretching of heteropolymers with uncorrelated random
sequences.  Introducing correlations into the 
sequence statistics will alter this local composition,
and may therefore affect stretching behavior strongly.
Here we consider three simple, prototypical forms of correlated
statistics.

When monomers of the same type are likely to be found within
a correlation ``length'' $\xi$ in the sequence, clusters of
like monomers occur with high probability.
The weight of necklace structures is clearly enhanced for such
blocky sequences, since regions of the chain with unfavorable 
solvation energy may be almost completely shielded from the solvent.
Specifically, the distribution of local sequence compositions is nearly 
binary and independent of region size $M$ for $M<\xi$:
\begin{equation}
p(q) \simeq {1\over 2}(\delta_{q,1}+\delta_{q,-1}).
\end{equation}
As a result, fluctuations in solvation energy of exposed
coil regions are of size $\Gamma \xi$.
If $\xi \sim N$, the free energy stabilizing necklace
structures is also macroscopic, ${\cal F}_{\rm rand} \sim N$.
With this macroscopic stabilization, the necklace phase 
will be stable over a finite range of $f$ even as
$N\rightarrow \infty$.

When correlations in monomer type decay algebraically,
rather than exponentially, similarly dramatic fluctuations
in local composition are possible.  Specifically, power law
correlations with decay exponent $\eta$ yield
$\langle (\delta q)^2 \rangle_q \sim M^{-\eta}$.
Because the sequence is one-dimensional, 
fluctuations are enhanced only for $\eta \leq 1$
At the crossover ($\eta=1$), 
$\langle (\delta q)^2 \rangle_q \sim M^{-1} \ln{M}$,
providing a weak stabilization of necklaces.
But as $\eta \rightarrow 0$, clusters of like monomers
may be arbitrarily large.  In this limit, 
necklace stabilization again becomes macroscopic.

Anticorrelations between like monomers, on the other hand,
tends to destabilize necklace structures.
In this case, the local composition is essentially ``neutral''
(i.e., $q\approx 0$) for regions larger than the scale of
anticorrelations, $\xi$.
For the statistics of neutrality fluctuations
in Coulombic systems, 
$\langle (\delta q)^2 \rangle_q \sim M^{-1/2}$,
fluctuations in the effective random potential,
$\Delta u = O(1)$, are especially small.
On scales larger than $\xi$, the chain is effectively
a homopolymer.
Consequently, necklace structures are not sufficiently stable 
to appear away from globule-coil coexistence.

These results suggest some basic principles
for designing mechanically robust 
heteropolymers.
Typical random sequences are poor candidates, since
they tend to form partially unfolded necklace structures
under strain.
Sequences in which solvophobic groups are heavily clustered
together will typically also permit stable ensembles of 
necklace structures.
Most promising are sequences whose correlations suppress 
fluctuations in local composition.
These may represent, for example, molecules with
widely distributed hydrophobic groups.
The compact native structures of such molecules
generally include important contacts linking
distant segments of the chain.
It is in part this topology 
imposed by nonlocal contacts that
provides a collective resistance to strain. 
Indeed, a common structural motif of
mechanical proteins, $\beta$-sheet secondary structure,
is typically rich in nonlocal hydrophobic contacts.
The elements of sequence statistics we predict to be favorable
for mechanical strength are thus related to topological features
of the native state suggested by computer 
simulations\cite{Schulten_PROT_99,Karplus_PNAS_00}.
It will be interesting to see how these basic
principles compare with simulations of evolutionary
design for mechanical strength.
For commonly used, coarse-grained models of proteins,
such simulations should be feasible using current 
computational resources and are currently underway in our
laboratory.

\section*{Acknowledgments}
We thank A. Yu. Grosberg and D. Klimov for fruitful
discussions and critical readings of this manuscript.
This work was supported by NIH.

\bibliographystyle{prsty}

\end{document}